\documentclass[aps,onecolumn,notitlepage,eqsecnum,nofootinbib,floatfix,superscriptaddress]{aastex62}
\usepackage{amsmath, amssymb, amsfonts, amsbsy,mathrsfs}

\usepackage{graphicx, epstopdf}
\DeclareGraphicsExtensions{.pdf}
\graphicspath{{./plots/}}

\usepackage[caption=false]{subfig}

\newcommand{\codename}{\mathtt{PoMiN}}

\begin{document}

\title{$\codename$: A Post-Minkowskian $N$-Body Solver}

\author{Justin Feng}
\affiliation{Theory Group, Department of Physics, University of Texas at Austin, USA}
\author{Mark Baumann}
\affiliation{Department of Physics, University of Texas at Austin, USA}
\author{Bryton Hall}
\affiliation{Department of Physics, University of Texas at Austin, USA}
\author{Joel Doss}
\affiliation{Department of Physics, University of Oregon, USA}
\author{Lucas Spencer}
\affiliation{Department of Physics, University of Texas at Austin, USA}
\author{Richard Matzner}
\affiliation{Theory Group, Department of Physics, University of Texas at Austin, USA}


\preprint{UTTG-02-18}

\begin{abstract}
In this paper, we introduce $\codename$, a lightweight $N$-body code based on the post-Minkowskian $N$-body Hamiltonian of Ledvinka et. al., which includes general relativistic effects up to first order in Newton's constant $G$, and all orders in the speed of light $c$. $\codename \> $  is written in $\mathtt{C}$ and uses a fourth-order Runge-Kutta integration scheme. $\codename$ has also been written to handle an arbitrary number of particles (both massive and massless), with a computational complexity that scales as $O(N^2)$. We describe the methods we used to simplify and organize the Hamiltonian, and the tests we performed (convergence, conservation, and analytical comparison tests) to validate the code.
\end{abstract}



\section{Introduction}

In general relativity (GR), the $N$-body problem (where $N \geq 2$) in an astrophysically realistic scenario requires the use of approximate and/or numerical methods \cite{PoissonWill}.  A widely used approximation to GR is the post-Newtonian (PN) formalism, which can be applied to weak gravitational fields and can be pushed to higher and higher orders (in powers of $G c^{-2}$ and $c^{-2}$) to create an increasingly accurate description of the field.  PN has proven to be very successful in several astrophysical applications of two- and three-body problems as well as for $N$-body problems with $N>3$, such as the solar system \cite{Kopeikin} and stars orbiting a supermassive black hole in a galactic nucleus \cite{Hamers,MikkolaMerritt}.

The PN formalism relies on the ``slow-motion'' assumption that characteristic velocities are lower than the speed of light.  If one wishes to relax this slow-motion condition (and obtain a ``fast-motion'' approximation), one may instead perform a post-Minkowskian (PM) approximation. Like the PN formalism, the PM formalism can be expressed in increasingly higher orders (this time, powers of Newton’s constant $G$) of increasing accuracy. The PM formalism has been used to model the wave zone around a coalescing binary black hole \cite{Blanchet} as well as in the effective one-body formalism of Damour for two-body systems \cite{EOB}.  Both of these use the PM formalism for an $N=2$ system.

The PN approximation with $N=2$ and $N \geq 3$ 
\cite{Futamase2007,Itoh2009,Schafer1987,LoustoNakano2008,Kupi2006,Bremetal2013,Aarseth2007,Will2013,Will2014}
(also see \cite{Blanchet} and references contained therein)
and the PM approximation with $N=2$ 
\cite{Blanchet,EOB,Westpfahl1979,Portilla1979,Portilla1980,Beletal1981,Westpfahl1985,Westpfahletal1987}, 
have been extensively studied in the literature, but relatively little has been done with the PM approximation for problems in which $N \geq 3$.  In this paper we describe $\codename$: a PM $N$-body solver for arbitrarily high $N$ that uses the fast-motion approximation to first order in $G$ and all orders in $c$.  $\codename$ is lightweight, meaning that it is small since it uses no external libraries (except $\mathtt{quadmath}$ if quadruple floating-point precision is desired) and is relatively simple, being a single file of fewer than 1000 lines written in $\mathtt{C}$.  As we show, $\codename$ has demonstrated the expected fourth-order convergence behavior (Section \ref{sec_convergence}), momentum conservation to machine precision, and good agreement with analytical values of momentum exchange (Section \ref{sec_analytical}).

Ledvinka, Sch{\"a}fer, and Bi{\v c}{\'a}k \cite{PM}, used a post-Minkowskian approximation to obtain a closed-form gravitational $N$-body Hamiltonian for nonspinning point particles\footnote{For another approach to the $N$-body problem in the PM approximation that can handle extended objects and spinning particles, see \cite{ZschockeSoffel}.} that takes into account general relativistic effects to first order in $G$ and all orders in $c$. The form of this $N$-body Hamiltonian, which we call the \textit{LSB Hamiltonian} (for the authors) is rather complicated, so it is appropriate to study the resulting dynamics computationally. 

$\codename$ uses a fourth-order Runge-Kutta (RK4) method to solve Hamilton's equations for the LSB Hamiltonian. We note that for long-term $N$-body simulations, one is often interested in integration methods that conserve energy (on average) over long time scales. Long-term $N$-body simulations therefore employ time-symmetric or symplectic integrators, which have improved energy conservation on long timescales compared to Runge-Kutta integrators \cite{Hutetal1997,Hutetal1995}.\footnote{We must also mention here the exactly conservative integrators of the type presented in \cite{Shadwicketal1995}.} Alhough it would be preferable to employ such methods in $\codename$, we have chosen to use an RK4 integrator for its relative simplicity and because it is explicit. While there exist explicit symplectic integrators for Hamiltonian systems that are simpler than RK4 (the leapfrog integrator, for instance), such integrators typically require separable Hamiltonians of the form $H(p,q)=T(p)+V(q)$ \cite{ForestRuth1990,*Yoshida1990}; the nonseparability of LSB Hamiltonian requires more sophisticated integrators, such as the partitioned-Runge-Kutta (PRK) methods (see Chapter 14 of \cite{SanzSernaCalvo1994} and references therein) or the splitting method of Tao \cite{Tao2016}. In spite of this current limitation (which we hope to resolve in future versions of $\codename$), we demonstrate that $\codename$ is suitable for studying problems in which gravitational interactions between objects are confined to short timescales (such as scattering problems). 

\section{Analytical Background}
%
%
\subsection{The Hamiltonian}
We briefly summarize the derivation of the LSB Hamiltonian; further details may be found in \cite{PM} and \cite{Schafer1986}. The starting point for the derivation of the LSB is the post-Minkowski Hamiltonian of \cite{Schafer1986}, which is a function of the coordinates and momenta for point particles, but is also a \textit{functional} of the gravitational field and its conjugate momenta. The derivation of the Hamiltonian in \cite{Schafer1986} begins with the ADM 3+1 split of the gravitational field \cite{MTW,ADM62,*ADM62b}, and the conversion of the boundary term in the ADM Hamiltonian to an integral over the bulk (done for a particular decomposition of the three-metric $g_{ij}$ and choice of gauge). Using the constraints, the result may be used to obtain a Hamiltonian that depends on the positions and conjugate momenta for point particle sources and the gravitational field $h^{TT}_{ij}$ and its conjugate momentum $\pi^{ij}_{TT}$, with $h^{TT}_{ij}$ being the transverse-traceless part of $h_{ij}:=g_{ij}-\delta_{ij}$ (the difference between the three-metric $g_{ij}$ and the Kronecker delta $\delta_{ij}$). 

The derivation of the LSB Hamiltonian involves solving the linearized field equations for $h^{TT}_{ij}$ with point particle sources, under the assumption that to first order in Newton's constant $G$, the field is generated entirely by unaccelerated particles. The solution $h^{TT}_{ij}(\textbf{x};\textbf{x}_a,\textbf{p}_a,\dot{\textbf{x}}_a)$ is a function of the spatial coordinate $\textbf{x}$, the positions of the particles $\textbf{x}_a$, their conjugate momenta $\textbf{p}_a$, and their velocities $\dot{\textbf{x}}_a$. To eliminate the fields from the Hamiltonian, a Routhian \cite{GoldsteinCM} is constructed from a Legendre transformation of the fields, so that it forms a Hamiltonian for the particles but a Lagrangian for the fields $h^{TT}_{ij}$. Upon noting that the aforementioned solutions $h^{TT}_{ij}(\textbf{x};\textbf{x}_a,\textbf{p}_a,\dot{\textbf{x}}_a)$ are nonradiative, and that the functional derivatives of the Routhian vanish on solutions of the field equations, one may substitute the solutions $h^{TT}_{ij}(\textbf{x};\textbf{x}_a,\textbf{p}_a,\dot{\textbf{x}}_a)$ without changing Hamilton's equations for the particles. Since the solutions $h^{TT}_{ij}(\textbf{x};\textbf{x}_a,\textbf{p}_a,\dot{\textbf{x}}_a)$ depend explicitly on the particle coordinates, their momenta, and the particle velocities, one can obtain a ``Hamiltonian'' for the particles that is independent of the fields. A coordinate transformation allows one to eliminate the dependence of the Hamiltonian on $\dot{\textbf{x}}_a$, and one obtains the LSB Hamiltonian \cite{PM}:
\begin{widetext}
\begin{equation} \label{SC2-LSBHamiltonian}
\begin{aligned} 
  H (\textbf{x}_a,\textbf{p}_a)= &\sum_{a= 1}^{N}\bar{m}_{a}- \frac{1}{2}G\sum_{a,b\neq a}^{N}\frac{\bar{m}_{a}\bar{m}_{b}}{r_{ab}}\left(1+ \frac{\textbf{p}^{2}_{a}}{\bar{m}^{2}_{a}}+ \frac{\textbf{p}^{2}_{b}}{\bar{m}^{2}_{b}}\right)+ \frac{1}{4}G\sum_{a,b\neq a}^{N} \biggl\{\frac{1}{r_{ab}}\left(7\mathbf{p}_{a}\cdot\mathbf{p}_{b}+ (\mathbf{p}_{a}\cdot\mathbf{n}_{ab})(\mathbf{p}_{b}\cdot\mathbf{n}_{ab})\right) \\
                     &\hspace{-1.75cm}-\frac{1}{r_{ab}}\frac{\left(\bar{m}_a\bar{m}_b\right)^{-1}}{\left(y_{ba}+1\right)^{2}y_{ba}}\biggl[ 2\left(2\left(\mathbf{p}_{a}\cdot\mathbf{p}_{b}\right)^{2}\left(\mathbf{p}_{b}\cdot\mathbf{n}_{ba}\right)^{2}- 2(\mathbf{p}_{a}\cdot\mathbf{n}_{ba})(\mathbf{p}_{b}\cdot\mathbf{n}_{ba})(\mathbf{p}_{a}\cdot\mathbf{p}_{b})\mathbf{p}^{2}_{b}+ \left(\mathbf{p}_{a}\cdot\mathbf{n}_{ba}\right)^{2}\mathbf{p}^{4}_{b} - \left(\mathbf{p}_{a}\cdot\mathbf{p}_{b}\right)^{2}\mathbf{p}^{2}_{b}\right)\frac{1}{\bar{m}^{2}_{b}}  \\
                     &\hspace{-.75cm} + 2\left(-\mathbf{p}^{2}_{a}\left(\mathbf{p}_{b}\cdot\mathbf{n}_{ba}\right)^{2}+ \left(\mathbf{p}_{a}\cdot\mathbf{n}_{ba}\right)^{2}\left(\mathbf{p}_{b}\cdot\mathbf{n}_{ba}\right)^{2}+ 2(\mathbf{p}_{a}\cdot\mathbf{n}_{ba})(\mathbf{p}_{b}\cdot\mathbf{n}_{ba})(\mathbf{p}_{a}\cdot\mathbf{p}_{b})+ \left(\mathbf{p}_{a}\cdot\mathbf{p}_{b}\right)^{2}- \left(\mathbf{p}_{a}\cdot\mathbf{n}_{ba}\right)^{2}\mathbf{p}^{2}_{b}\right)  \\
                    & \hspace{-.75cm} + \left(-3\mathbf{p}^{2}_a\left(\mathbf{p}_{b}\cdot\mathbf{n}_{ba}\right)^{2}+ \left(\mathbf{p}_{a}\cdot\mathbf{n}_{ba}\right)^{2}\left(\mathbf{p}_{b}\cdot\mathbf{n}_{ba}\right)^{2}+ 8(\mathbf{p}_{a}\cdot\mathbf{n}_{ba})(\mathbf{p}_{b}\cdot\mathbf{n}_{ba})(\mathbf{p}_{a}\cdot\mathbf{p}_{b})+ \mathbf{p}^{2}_{a}\mathbf{p}^{2}_{b}-3\left(\mathbf{p}_{a}\cdot\mathbf{n}_{ba}\right)^{2}\mathbf{p}^{2}_{b}\right)y_{ba}\biggr]\biggr\}
\end{aligned}
\end{equation}
\end{widetext}

\noindent where $\textbf{x}_a$ and $\textbf{p}_a$ are the respective positions and conjugate momenta of the particles, and the following quantities are defined:
\begin{equation} \label{SC2-LSBHamiltonianDefns}
\begin{aligned} 
\bar{m}_a &:=\sqrt{m_a^2+\textbf{p}_a^2}\\
r_{ab}&:=|\textbf{x}_a - \textbf{x}_b|=\sqrt{\left(\textbf{x}_a - \textbf{x}_b\right)\cdot\left(\textbf{x}_a - \textbf{x}_b\right)}\\
\textbf{n}_{ab} &:=r_{ab}^{-1}\,(\textbf{x}_a - \textbf{x}_b)\\
y_{ba}&:=\bar{m}_b^{-1}\,\sqrt{m^{2}_{b}+\left(\mathbf{n}_{ba} \cdot \mathbf{p}_{b}\right)^{2}}\\
\end{aligned}
\end{equation}

\noindent with $m_a$ being the rest mass, which is not to be confused with $\bar{m}_a$, which one may recognize as the relativistic kinetic energy of a particle in flat spacetime. One may also recognize $r_{ab}$ and $\textbf{n}_{ab}$ as the respective distance of separation between particles $a$ and $b$ and the unit separation vector $\textbf{n}_{ab}$.

%
%
\subsection{Splitting the Hamiltonian}
In spite of its simplicity, the LSB Hamiltonian (\ref{SC2-LSBHamiltonian}) is still rather complicated; to simplify the Hamiltonian, we introduce the following scalar quantities:
\begin{equation} \label{ThetaXi}
\begin{aligned}
\Theta_{ab} &:= \mathbf{p}_{a}\cdot\mathbf{n}_{ba}\\
\Xi_{ab} &:= \mathbf{p}_{a}\cdot\mathbf{p}_{b}\\
\end{aligned}
\end{equation}

\noindent We decompose the Hamiltonian (\ref{SC2-LSBHamiltonian})
into three parts:
\begin{equation} \label{HamiltonianDecomp}
H= H_{1}+ H_{2}+ H_{3}
\end{equation}

\noindent where we define the following:
\begin{equation} \label{HamiltonianPart1}
  H_{1} := \sum_{a= 1}^{N}\bar{m}_{a}- \frac{1}{2}G\sum_{a,b\neq a}^{N}\frac{\bar{m}_{a}\bar{m}_{b}}{r_{ab}}\left(1+ \frac{\textbf{p}^{2}_{a}}{\bar{m}^{2}_{a}}+ \frac{\textbf{p}^{2}_{b}}{\bar{m}^{2}_{b}}\right)
\end{equation}
\begin{equation} \label{HamiltonianPart2}
  H_{2} := \frac{1}{4}G\sum_{a,b\neq a}^{N}\frac{1}{r_{ab}}\left(7 \, \Xi_{ab}- \Theta_{ab} \, \Theta_{ba}\right)
\end{equation}

\begin{widetext}
\begin{equation} \label{HamiltonianPart3}
\begin{aligned}
  H_{3} := -\frac{1}{4}G\sum_{a,b\neq a}^{N}\frac{1}{r_{ab}}\frac{\left(\bar{m}_a\bar{m}_b\right)^{-1}}{\left(y_{ba}+1\right)^{2}y_{ba}}\biggl[&2\left[-2 \, \Xi^{2}_{ab} \, \Theta^{2}_{ba}+ 2 \,\Theta_{ab} \, \Theta_{ba}\Xi_{ab} \, \mathbf{p}^{2}_{b}+ \Theta^{2}_{ab} \, \mathbf{p}^{4}_{b}- \Xi^{2}_{ab} \, \mathbf{p}^{2}_{b}\right]\frac{1}{\bar{m}^{2}_{b}} \\
         &+2\left[\mathbf{p}^{2}_{a} \, \Theta^{2}_{ba}- \Theta^{2}_{ab} \, \Theta^{2}_{ba}- 2 \, \Theta_{ab} \, \Theta_{ba} \, \Xi_{ab}+ \Xi^{2}_{ab}- \Theta^{2}_{ab} \, \mathbf{p}^{2}_{b}\right] \\
         &+\left[3\, \mathbf{p}^{2}_a \, \Theta^{2}_{ba}- \Theta^{2}_{ab} \, \Theta^{2}_{ba}- 8 \, \Theta_{ab} \, \Theta_{ba} \, \Xi_{ab}+ \mathbf{p}^{2}_{a} \, \mathbf{p}^{2}_{b}-3 \, \Theta^{2}_{ab} \, \mathbf{p}^{2}_{b}\right]y_{ba}\biggr]. \end{aligned}
\end{equation}
\end{widetext}

\noindent The first term in $H_1$ is the sum of the special-relativistic energies $\bar{m}_a$ for each particle, and the second term in $H_1$ contains the Newtonian gravitational potential. $H_2$ and $H_3$ contain post-Minkowskian contributions to the Hamiltonian. Note that the individual terms in $H_{1}$ and $H_{2}$ are symmetric, but those in $H_{3}$ are not due to the $y_{ba}$ factors. In principle, $H_{3}$ may be rewritten in a symmetric form due to the summation over $a$ and $b$.

A difficulty that one encounters is the fact that $y_{ba}$ may vanish when the particle labeled by $b$ is massless ($m_b=0$) and the momentum $\textbf{p}_b$ is orthogonal to the separation vector $\textbf{n}_{ba}$. Naively, one might expect that since $H_3$ contains a factor of $y_{ba}^{-1}$, both $H_3$ and its derivatives will diverge as a result. It turns out that when $m_b=0$, both $H_3$ and its derivatives simplify, and the problematic factor $y_{ba}^{-1}$ is canceled by factors of $\Theta_{ba}$ in the numerator. We briefly describe what happens for the case of $H_3$.  First, we note that if $m_{b}= 0$, then $\bar{m}_b^2=\textbf{p}_b^2$ and $\Theta_{ba}= \sigma y_{ba}|\mathbf{p}_{b}|$, where
$\sigma= \mathrm{sgn}(\Theta_{ba})$. If we replace $\bar{m}_b^2$ and $\Theta_{ba}$ accordingly, some terms cancel, and the quantity in the large square brackets of Equation (\ref{HamiltonianPart3}) becomes proportional to $y_{ab}$. The resulting expression for $H_3$ in the limit $m_b \rightarrow 0$ is:
\begin{widetext}
\begin{equation} \label{HamiltonianPart3MasslessLimit}
\begin{aligned}
  \lim_{m_b \rightarrow 0} H_{3}= -\frac{1}{4}G\sum_{a,b\neq a}^{N}\frac{1}{r_{ab}}\frac{\left(\bar{m}_a\bar{m}_b\right)^{-1}}{\left(y_{ba}+1\right)^{2}} \biggl[ & 4y_{ba}\Xi^{2}_{ab}- 2y_{ba}\mathbf{p}^{2}_{a}\mathbf{p}^{2}_{b}+ 2y_{ba}\Theta^{2}_{ab}\mathbf{p}^{2}_{b} -3y^2_{ba}\mathbf{p}^{2}_{a}\mathbf{p}^{2}_{b}+ y_{ba}\Theta^{2}_{ab}\mathbf{p}^{2}_{b}- 8\Theta_{ab}\Theta_{ba}\Xi_{ab} + \mathbf{p}^{2}_{a}\mathbf{p}^{2}_{b} \\
  & - 3\Theta^{2}_{ab}\mathbf{p}^{2}_{b} \biggr].
\end{aligned}
\end{equation}
\end{widetext}

\noindent One may be tempted to use the above expression (\ref{HamiltonianPart3MasslessLimit}) as the Hamiltonian for massless particles; this works when \textit{all} particles in the system are massless, but one cannot use terms of the form in Equation (\ref{HamiltonianPart3MasslessLimit}) to describe the interaction between massive and massless particles. This is because derivatives of the Hamiltonian and the limit $m_b \rightarrow 0$ do not commute; one obtains different results for the derivatives of the Hamiltonian depending on whether the limit is taken before or after the derivatives are performed. The correct procedure is to take the limit $m_b \rightarrow 0$ \textit{after} taking the derivatives of the Hamiltonian.

%
%
\subsection{Formal Derivative Methods}
In this section, we describe a method for systematically computing the derivatives of the LSB Hamiltonian, in which we decompose the derivatives of the Hamiltonian according to the chain rule. The equations of motion for a Hamiltonian system are described by
Hamilton's equations:
\begin{equation}
\begin{aligned} \label{HamiltonEquations}
  \dot{p}_{i}^a\equiv -\frac{\partial H}{\partial q^{i}_a} && \dot{q}^{i}_a\equiv \frac{\partial H}{\partial p_{i}^a},
\end{aligned}
\end{equation}
where $q^{i}_a$ is the $i\mathrm{th}$ component of $\mathbf{q}_{a}$ and $p_{i}^a$ is the $i\mathrm{th}$ component of $\mathbf{p}_{a}$. To compute the equations of motion for all $N$ particles, one can use the formal derivative of a Hamiltonian, $\frac{\partial H}{\partial z}$. If the Hamiltonian is a function of scalar quantities $\Phi^{A}(z)$, then $\frac{\partial H}{\partial z}$ is a function of $\Phi^{A}(z)$ and $\frac{\partial(\Phi^{A})}{\partial z}$, such that
\begin{equation}
\begin{aligned}
  H&= H\left(\Phi^{A}(z)\right) \\
  \frac{\partial H}{\partial z}&= F_{(z)}\left(\Phi^{A},\frac{\partial \Phi^{A}}{\partial z}\right),
\end{aligned}
\end{equation}
where $F_{(z)}$ is given by the following expression (this is just the chain rule applied to the derivative of the Hamiltonian):
\begin{equation} \label{eqn:10}
  F_{(z)}\left(\Phi^{A},\frac{\partial \Phi^{A}}{\partial z}\right) = \sum_{A}\frac{\partial H}{\partial(\Phi^{A})}\frac{\partial \Phi^{A}}{\partial z}.
\end{equation}
For the Hamiltonian defined in Equations (\ref{HamiltonianDecomp})--(\ref{HamiltonianPart3}), $\Phi^A$ represents the following set:
\begin{equation} \label{PhiSet}
\Phi^{A}\in \left\{\bar{m}_{a},\mathbf{p}^{2}_{a},r_{ab},y_{ba},\Theta_{ab},\Xi_{ab}\right\}.
\end{equation}

For a system of $N$ particles, this is a set of $2N+ 4(N^{2}- N)= 4N^{2}- 2 N$ scalar quantities. Thus, the formal derivative $\frac{\partial \Phi^{A} }{\partial z}$ represents the set
\begin{equation} \label{PhiDerivSet}
\begin{aligned}
  \frac{\partial \Phi^{A}}{\partial z}\in \biggl\{&\frac{\partial\bar{m}_{a}}{\partial q^i_c},\frac{\partial\mathbf{p}^{2}_{a}}{\partial q^i_c},\frac{\partial r_{ab}}{\partial q^i_c},\frac{\partial y_{ba}}{\partial q^i_c},\frac{\partial\Theta_{ab}}{\partial q^i_c},\frac{\partial\Xi_{ab}}{\partial q^i_c}; \\
 & \frac{\partial\bar{m}_{a}}{\partial p_i^c},\frac{\partial\mathbf{p}^{2}_{a}}{\partial p_i^c},\frac{\partial r_{ab}}{\partial p_i^c},\frac{\partial y_{ba}}{\partial p_i^c},\frac{\partial\Theta_{ab}}{\partial p_i^c},\frac{\partial\Xi_{ab}}{\partial p_i^c}\biggr\}.
\end{aligned}
\end{equation}

Each particle in this system exists in a six-dimensional phase space, so every scalar quantity has $6N$ derivatives. As such, the total number of derivatives for this system is $6N\times (4N^{2}- 2 N)= 24N^{3}- 12N^{2}$. While this suggests that the number of calculations scales as $\sim N^{3}$, the complexity is reduced by computing only nonvanishing derivatives. The nonvanishing derivatives are those in the set (\ref{PhiDerivSet}) where the particle label $c$ matches that of either $a$ or $b$, such that $c\in \{a,b\}$, reducing the $6N$ derivatives to $6\times 2= 12$. The total number of derivatives is now $12\times (4N^{2}-  2N)= 48N^{2}- 24N$. Therefore, the total number of quantities $\Phi^{A}(z)$ and $\frac{\partial \Phi^{A}}{\partial z}$ that must be calculated at every timestep is $52N^{2}- 26N$. The complexity can be further reduced by observing that the following derivatives always vanish:
\begin{equation}
\begin{aligned} 
  \frac{\partial(\mathbf{p}^{2}_{a})}{\partial q^i_c} &= 0;\\
   \frac{\partial r_{ab}}{\partial p_i^c} &= 0; \\
  \frac{\partial\bar{m}_{a}}{\partial q^i_c}&= 0; \\
   \frac{\partial\Xi_{ab}}{\partial q^i_c}&= 0.
\end{aligned}
\end{equation}

\noindent Hamilton's equations (\ref{HamiltonEquations}) can now be rewritten in terms of the scalar quantities $\Phi^{A}(z)$ and $\frac{\partial \Phi^{A}}{\partial z}$:
\begin{equation} \label{HamiltonEquations2}
\begin{aligned}
  \dot{p}_i^c&\equiv -\frac{\partial H}{\partial q^i_c}= -F_{(q^i_c)}\left(\Phi^{A},\frac{\partial \Phi^{A}}{\partial q^i_c}\right) \\
  \dot{q}^{ic}&\equiv \frac{\partial H}{\partial p_i^c}= F_{(p_i^c)}\left(\Phi^{A},\frac{\partial \Phi^{A}}{\partial p_i^c}\right),
\end{aligned}
\end{equation}
where $F_{(z)}$ is the formal derivative of the $H$ defined in equations (\ref{HamiltonianDecomp})--(\ref{HamiltonianPart3}). We compute an expression for the formal derivatives\footnote{We take the limit $m_b \rightarrow 0$ in the derivatives of the Hamiltonian, which allows us to get rid of factors of $y_{ba}$ in the denominator of Equation (\ref{HamiltonEquations2}).} $F_{(z)}$ of the Hamiltonian using a computer algebra system (we use \textit{Mathematica}) and convert the resulting expression for $F_{(z)}$ to the $\mathtt{C}$ programming language by means of character and string replacements. The right-hand side of Hamilton's equations (\ref{HamiltonEquations2}) is evaluated by first computing $\Phi^A$ and its derivatives, then inserting the result into the $\mathtt{C}$ implementation of the function $F_{(z)}$.

Before proceeding, we mention that in addition to their organizational appeal, the formal derivative methods we have described in this section naturally lead to more efficient codes for a (generic) complicated Hamiltonian. This is because a code based on the formal derivative methods only requires computing the derivatives $\frac{\partial \Phi^{A}}{\partial z}$ once in each timestep; the numerical implementation of the ``brute force'' method will (barring some extraordinary algebraic simplification) in general require performing computations corresponding to the derivatives $\frac{\partial \Phi^{A}}{\partial z}$ more than once. Of course, our comments here are generic and are intended for problems involving a complicated Hamiltonian (such as the LSB Hamiltonian) for which there is no obvious simplification, either in the Hamiltonian itself or in the resulting Hamilton's equations.

\section{Numerical Methods and Tests}
\subsection{Overview of Numerical Methods}
\label{sec_num_methods}
$\codename$ is open source and is designed with simplicity in mind.\footnote{The code is written in $\mathtt{C}$ and the source consists of a single file. The source code contains three primary functions, one of which is the $\mathtt{main}$ computational loop.  The other two functions, named $\mathtt{HamiltonEquations}$ and $\mathtt{DHamiltonian}$, are used to compute the right-hand side of Hamilton's equations (\ref{HamiltonEquations2}); in particular, $\mathtt{HamiltonEquations}$ computes the quantities $\Phi^A$ and calls $\mathtt{DHamiltonian}$, which computes the derivatives of the Hamiltonian defined in Equations (\ref{HamiltonianDecomp}--\ref{HamiltonianPart3}).} $\codename$ uses an RK4 integration method to numerically solve Hamilton's equations (\ref{HamiltonEquations2}) as given by the Hamiltonian defined in equations (\ref{HamiltonianDecomp}--\ref{HamiltonianPart3}). We employ a simple global adaptive time-stepping scheme\footnote{The user has the option to disable this feature.} based on the Courant---Friedrichs---Lewy (CFL) condition~\cite{Courant}. An adaptive time-stepping scheme is particularly useful for scattering problems; in scattering problems, particles start out with a large separation distance, but the distance of closest approach may be several orders of magnitude smaller than the initial separation distances. Such a large initial separation is necessary in scattering problems because it is difficult to invert Hamilton's equations to solve for the full dependence of the conjugate momenta $\textbf{p}_a$ on the particle velocities $\dot{\textbf{x}}_a$, except in the limit where the particles are separated by large distances, in which case the canonical momenta are well approximated by the special-relativistic momenta. Our implementation of the CFL condition places a limit on the ratio between the distance a particle moves in a single timestep and the distance to its nearest neighbor. Given a timestep $\Delta t$, the change in the magnitude of the relative particle distance is $\Delta r_{ab}\simeq |\mathbf{v}_{ab}|\Delta t$, where $|\mathbf{v}_{ab}|\equiv \sqrt{\dot{\mathbf{q}}_{a} - \dot{\mathbf{q}}_{b}}$. Thus, our CFL condition is
\begin{equation} \label{CFL-Condition}
  \frac{\Delta r_{ab}}{r_{ab}}= \frac{|\mathbf{v}_{ab}|\Delta t}{r_{ab}} \leq C,
\end{equation}

\noindent where $C$ is the Courant number, the upper limit for this ratio at a given
timestep. For the adaptive time-stepping algorithm, this condition must
hold for the pair of particles that are closest to each other. If
this condition fails, the timestep is recomputed using the following formula:
\begin{equation} \label{AdaptiveTimestepFormula}
  \Delta t \equiv C \frac{r_{ab}}{|\mathbf{v}_{ab}|}.
\end{equation}

%
%
\subsection{Convergence Tests}
\label{sec_convergence}
We perform implicit self-consistent convergence tests to check that our code behaves in a manner expected of a fourth-order code. Such convergence tests are particularly important for situations in which there is no exact analytical solution available for comparison, and can reveal the presence of mistakes and bugs in the code~\cite{BaumgarteShapiro}\footnote{It is worth reiterating the admonishment found at the end of chapter 9 in \cite{Alcubierre}: one should not trust any numerical calculation for which no convergence tests have been performed.} The convergence tests we have performed are based on Richardson extrapolation, in which one conjectures that the numerical result\footnote{One might, for instance, use the values of the phase space coordinates $z$ at the final timestep.} for any phase space coordinate $z$ differs from the analytical result by a power series expansion of the timestep \cite{Richardson1910,Alcubierre,BaumgarteShapiro,Choptuik}:
\begin{equation} \label{eqn:16}
  z_{nm}(h)= z_{an}+ e_{1}h+ e_{2}h^{2}+ e_{3}h^{3}+ e_{4}h^{4}+\cdots 
\end{equation}

\noindent Since RK4 is a fourth-order integration method, the expected errors in the numerical result should be of the order of $h^{4}$, such that
$e_{1}= e_{2}= e_{3}= 0$. Following Choptuik~\cite{Choptuik}, we define a convergence factor:
\begin{equation} \label{eqn:17}
  Q= \left|\frac{z_{nm}(4h)- z_{nm}(2h)}{z_{nm}(2h)- z_{nm}(h)}\right|,
\end{equation}

\noindent where the use of absolute values denotes the Euclidean norm. Using Equation~\ref{eqn:17}, one can calculate the convergence factor $Q$ as the timestep $h$ is repeatedly halved in the limit $ h \rightarrow 0$:
\begin{equation} \label{QFactor}
\begin{aligned}
  \lim_{h \rightarrow 0}Q&= \lim_{h \rightarrow 0}\left|\frac{z_{nm}(h)- z_{nm}(\frac{h}{2})}{z_{nm}(\frac{h}{2}))- z_{nm}(\frac{h}{4}))}\right|  \\
  &\simeq \lim_{h \rightarrow 0}\left|\frac{(e_{4}h^{4}+ {e_5}{O}(h^5))- (e_{4}(\frac{h}{2})^{4}+ O(h^5))}{(e_{4}(\frac{h}{2})^{4}+ O(h^5))- (e_{4}(\frac{h}{4})^{4}+ O(h^5))}\right|  \\
&\simeq \lim_{h \rightarrow 0}\left|\frac{e_{4}h^{4}- e_{4}(\frac{h}{2})^{4}}{e_{4}(\frac{h}{2})^{4}- e_{4}(\frac{h}{4})^{4}}\right|= \lim_{h \rightarrow 0}\left|\frac{1- (\frac{1}{2})^{4}}{(\frac{1}{2})^{4}- (\frac{1}{4})^{4}}\right| \\
&= \lim_{h \rightarrow 0}\left|\frac{1- \frac{1}{16}}{\frac{1}{16}- \frac{1}{256}}\right|\\
&= 16 
\end{aligned}
\end{equation}

\noindent The above computation (\ref{QFactor}) demonstrates that in the limit $h\rightarrow 0$, the convergence factor $Q$ should converge to a value of 16 for a code based on the RK4 method.

\begin{figure*}
 \includegraphics[width=\linewidth]{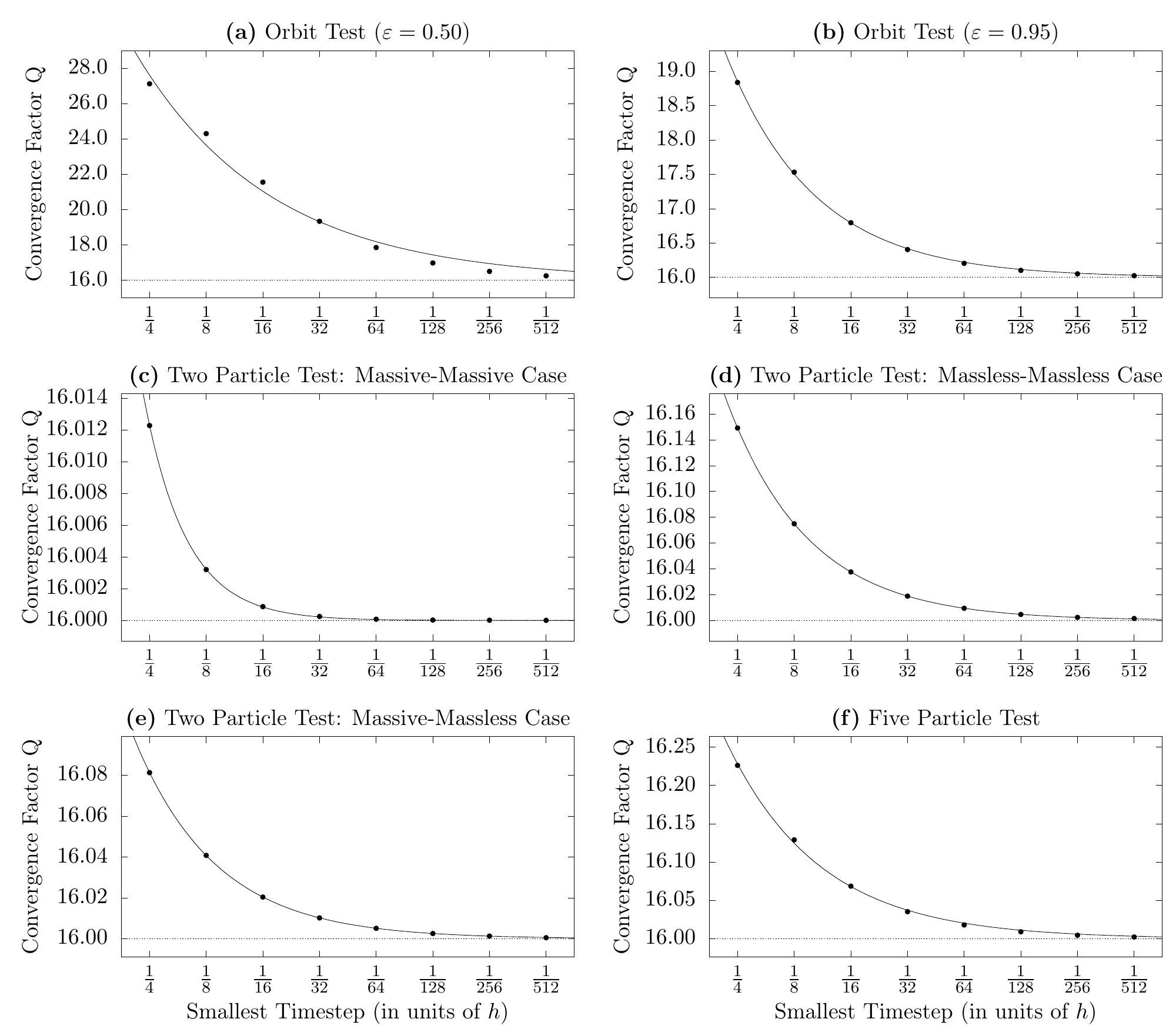}
\caption{Results of six convergence tests . Here, computations are performed to quadruple floating-point precision. In each plot, we have plotted (using dots) the convergence factor $Q$ (for the quantity $\textbf{p}^2$ of a particle in the system) against the smallest timestep used to compute $Q$. The timestep are given in units of $h$, where $h$ is the largest timestep used in the convergence test. For each plot, the values of $h$ (in natural units $c=G=1$) are as follows: for (a), $h=6.15 \times 10^{13}$; for (b), $h=2.70 \times 10^{13}$; and for plots (c) through (f), $h=0.05$. In plot (a), the initial number of timesteps is $200$, in plot (b), the initial number of timesteps is 1000, and in plots (c)-(f), the initial number of timesteps is $5$. We have performed a fit to an exponential curve that we have included in the plots.}
\label{FIG-ConvergenceTests}
\end{figure*}

In Figure \ref{FIG-ConvergenceTests}, we present the results of six convergence tests of our code\footnote{To compute $Q$ for very small $h$, we pushed our convergence tests beyond double floating-point precision and used the $\mathtt{quadmath}$ library.  This feature of the code can be turned off and on; in the case where more precision is needed, quadruple floating-point precision can be turned on, and in the case where that level of precision is not needed and a lighter-weight code with more speed is desired, double floating-point precision can be turned on.}. In each case, we examine the quantity $\textbf{p}^2$ for the particle of lower mass or zero mass, with the exception of Figure \ref{FIG-ConvergenceTests}(f), which we shall discuss later. The first three tests (Figures \ref{FIG-ConvergenceTests}(a)-(c)) each involved two massive particles, with one particle being much more massive than the other. In Figures \ref{FIG-ConvergenceTests}(a) and \ref{FIG-ConvergenceTests}(b), initial conditions were chosen so that the particles follow orbits of high eccentricity; in Figure \ref{FIG-ConvergenceTests}(a) we chose an eccentricity of $\varepsilon=0.50$, and in Figure \ref{FIG-ConvergenceTests}(b), we chose an eccentricity of $\varepsilon=0.95$. In both Figures \ref{FIG-ConvergenceTests}(a) and \ref{FIG-ConvergenceTests}(b), the central mass was chosen to be unity (in natural units $c=G=1$), and the lighter mass was chosen to have a value of $1.66 \times 10^{-7}$.

Figures \ref{FIG-ConvergenceTests}(c) through \ref{FIG-ConvergenceTests}(e) describe the convergence tests for the scattering of two particles. In all cases, the impact parameter is $19.61$ (again, in natural units $c=G=1$), the total rest energy is unity, and the initial conditions are chosen near the point of closest approach. Figure \ref{FIG-ConvergenceTests}(c) involves the ultrarelativistic scattering of two massive particles of unequal mass, with the higher mass having a value of $m_1=0.0498$, and the lower mass having a value of $m_2= \pi/4 m_1$, and each particle having a momentum of magnitude $|\textbf{p}|=10 \, m_1$ at late times. Figure \ref{FIG-ConvergenceTests}(d) involves the scattering of two massless particles; here, the system is symmetric, and the magnitude of the late-time momenta for each particle has a value of $0.50$. Figure \ref{FIG-ConvergenceTests}(e) involves the scattering between a massive and a massless particle; the mass of the massive particle has a value of $0.541$, and each particle has a momentum of magnitude $0.354$ in the late-time limit.
 
Figure \ref{FIG-ConvergenceTests}(f) describes the convergence test for a five-particle system involving two massive particles and three massless particles. The initial conditions for this system were formed from the initial conditions for the massless-massive scattering test in Figure \ref{FIG-ConvergenceTests}(e) and the massless-massless scattering test in \ref{FIG-ConvergenceTests}(d), with the addition of a stationary massive particle of unit mass at the origin. The convergence test was performed for the quantity $\textbf{p}^2$ on the lighter massive particle, which has a mass of $0.541$. Since this five-particle test is the most comprehensive convergence test for $\codename$, we also list the values of the convergence factor $Q$ for this test in Table \ref{TAB-ConvergenceFive}.

In all cases, the convergence factor $Q$ monotonically converges to the value $16$; in most cases (Figures \ref{FIG-ConvergenceTests}(b) through \ref{FIG-ConvergenceTests}(f) in particular), $Q$ exhibits exponential convergence to $16$. The convergence test results summarized in Figure \ref{FIG-ConvergenceTests} and Table \ref{TAB-ConvergenceFive} demonstrate that $\codename$ converges in a manner that one expects of a fourth-order code.

\begin{table} 
\centering
\begin{tabular}{|c|c|}
\hline
$\>$ Smallest Timestep $\>$ & Convergence Factor $Q$\\
\hline
${h}/{4}$  &  $\>$ $16.2262$ $\>$  \\
\hline
${h}/{8}$  &  $\>$ $16.1291$ $\>$  \\
\hline
${h}/{16}$  &  $\>$ $16.0686$ $\>$  \\
\hline
${h}/{32}$  &  $\>$ $16.0353$ $\>$  \\
\hline
${h}/{64}$  &  $\>$ $16.0179$ $\>$  \\
\hline
${h}/{128}$  &  $\>$ $16.0090$ $\>$  \\
\hline
${h}/{256}$  &  $\>$ $16.0045$ $\>$  \\
\hline
${h}/{512}$  &  $\>$ $16.0023$ $\>$  \\
\hline
\end{tabular}
\caption{This table lists the value of the convergence factor $Q$ for $\textbf{p}^2$ lighter massive particle in the Five-particle test of Figure \ref{FIG-ConvergenceTests}(f), with respect to the smallest timestep (here, $h=0.05$) used to compute the convergence factor $Q$.} \label{TAB-ConvergenceFive}
\end{table}

%
%
\subsection{Analytical Momentum Exchange and Conservation of Momentum Tests}
\label{sec_analytical}

To ensure that our code is indeed modeling the system described by the LSB Hamiltonian (\ref{SC2-LSBHamiltonian}), it is important to have analytical results to compare with, in particular, those that include effects beyond that of Newtonian gravity. However, the complexity of the LSB Hamiltonian (\ref{SC2-LSBHamiltonian}) and the resulting Hamilton equations \ref{HamiltonEquations} limits the analytical results available for comparison. Fortunately, \cite{PM} present the following (approximate) momentum exchange formula for scattering problems:
\begin{equation}
\begin{aligned} \label{delta_p}
\Delta \textbf{p} =& -2\frac{{\textbf{b}}}{{\textbf{b}}^2} \frac{G}{p}
\frac{\bar{m}_1^2 \bar{m}_2^2}{\bar{m}_1 + \bar{m}_2 } \\
& \times
\biggl[
1+\left(\frac{1}{\bar{m}_1^2}+\frac{1}{\bar{m}_2^2}+\frac{4}{\bar{m}_1 \bar{m}_2} \right){p}^2 + \frac{{p}^4}{\bar{m}_1^2 \bar{m}_2^2 }
\biggr]~.
\end{aligned}
\end{equation}

This formula is written in the center-of-mass frame, where both particles have momenta of magnitude $p$, and is derived assuming that the spatial trajectory of each particle approximates a straight line that is (anti-) parallel to the trajectory of the other particle. Equation (\ref{delta_p}) is therefore valid only when the angle of deflection is small, or when the impact parameter $|\textbf{b}|$ is large. The vector $\textbf{b}$ points in the direction of the perpendicular separation between the trajectories, with a magnitude given by the impact parameter $|\textbf{b}|$. The result $\Delta \textbf{p}$ describes the total momentum exchange between two particles during the scattering process.

\begin{figure*}
\subfloat[\label{FIG-ScTMixed}]{%
  \includegraphics[width=\linewidth]{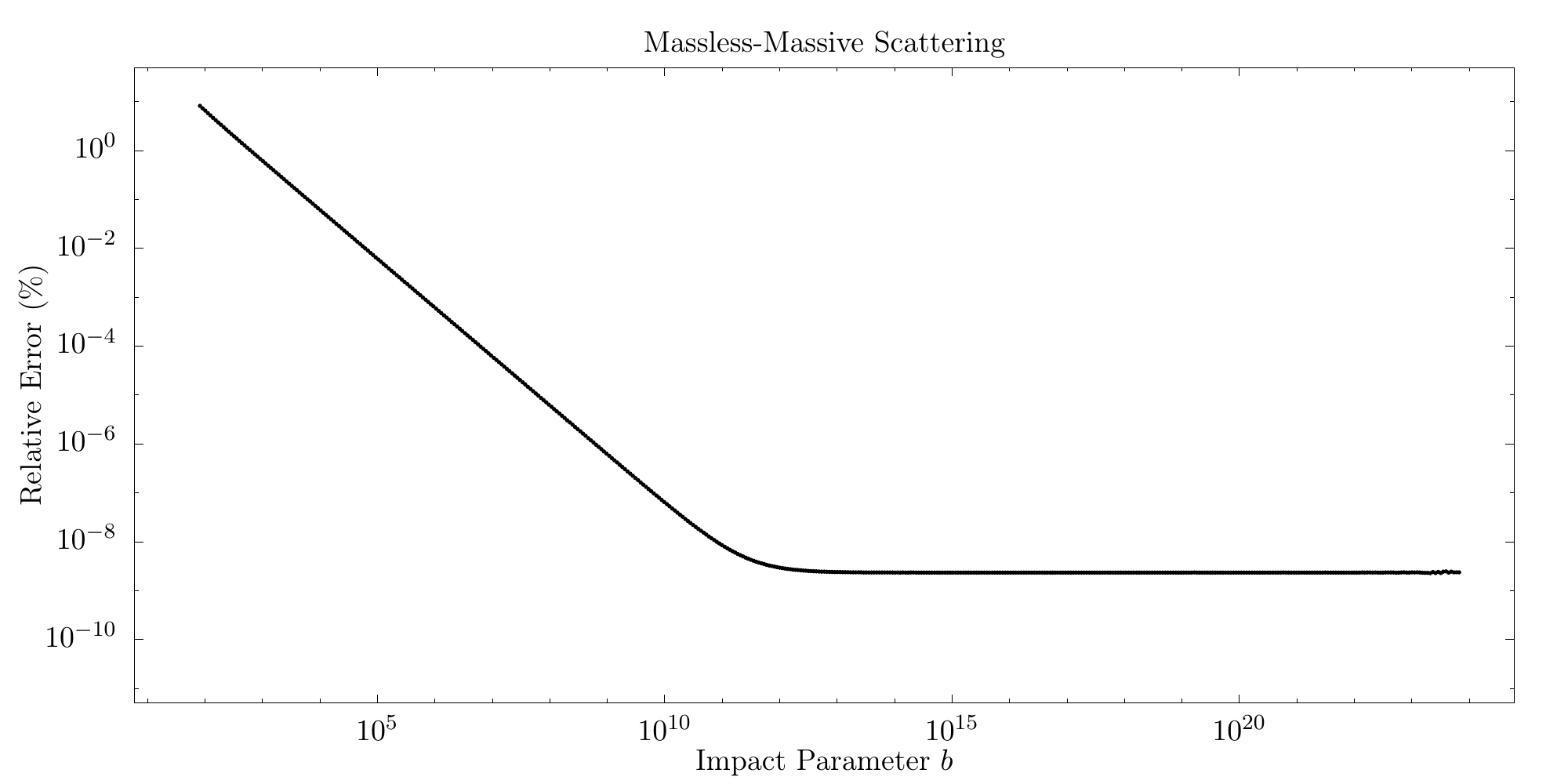}%
}

\subfloat[\label{FIG-ScTMassless}]{%
  \includegraphics[width=.49\linewidth]{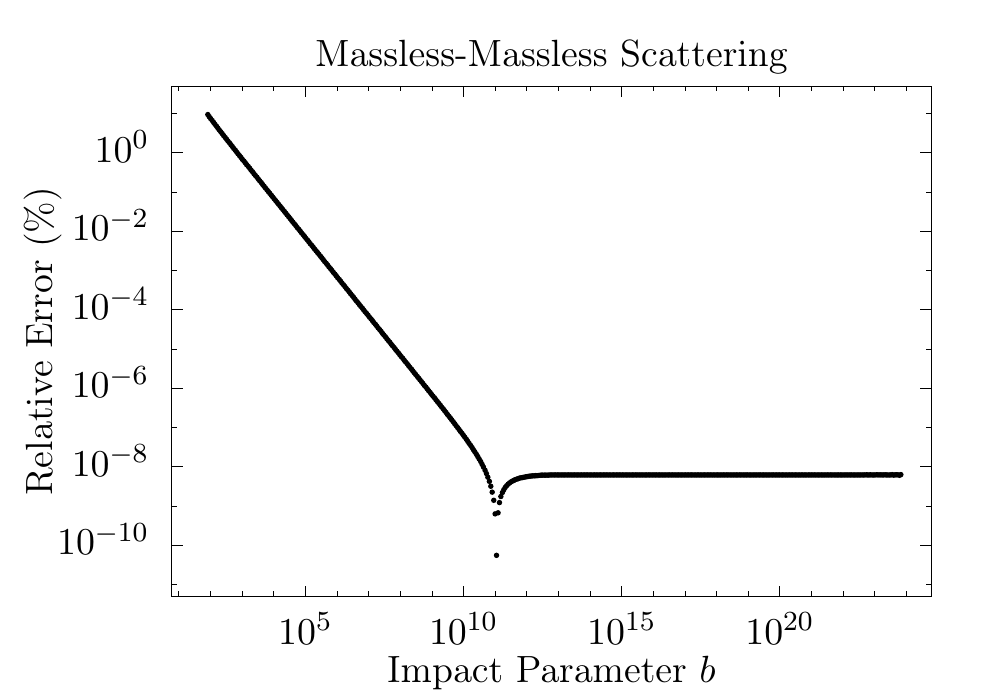}%
}\hfill
\subfloat[\label{FIG-ScTMassive}]{%
  \includegraphics[width=.49\linewidth]{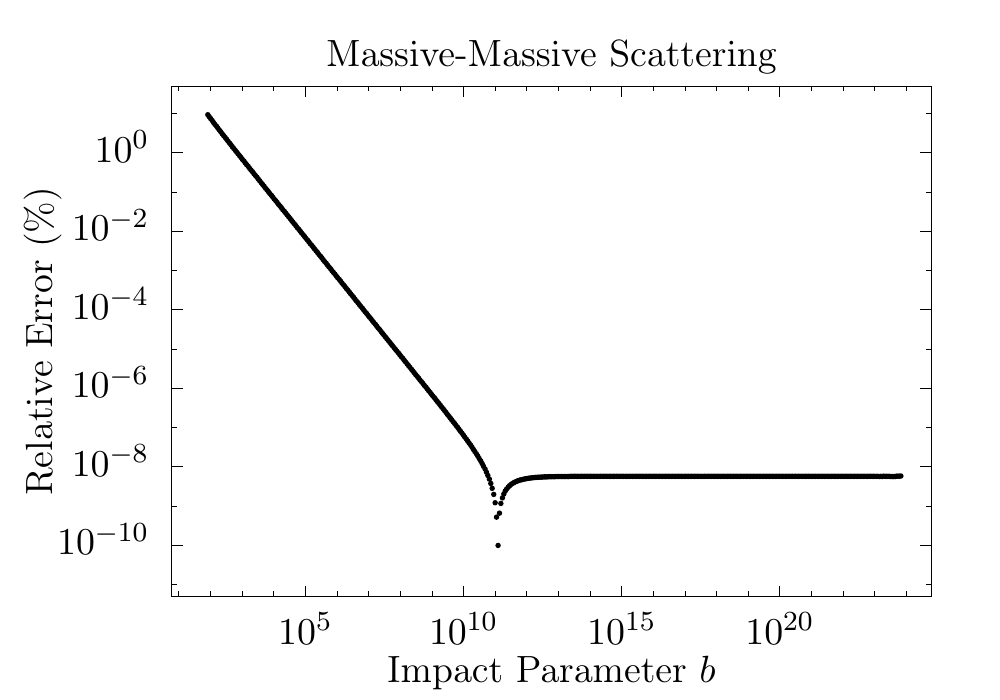}%
}
\caption{Plots of the relative error, Equation (\ref{relerror}), between the momentum exchange $\Delta p_{num}$ as computed by $\codename$ and the analytical momentum exchange $\Delta p_{an}$, as a function of the impact parameter $b$. In each case, the initial conditions were chosen so that the initial trajectories of the particles are antiparallel and offset by the impact parameter $b$, with an initial separation distance (in the direction of motion) of $10^5 b$. Moreover, the initial momenta and masses were chosen so that the total mass-energy is $1$ (we work in the natural units $G=1$, and $c=1$). In the massive-massive scattering case (Figure \ref{FIG-ScTMassive}), we used a mass ratio of $\pi/4$; $m_2= \pi / 4 \, m_1$.}
\label{FIG-ScatteringTestsRelativeError}
\end{figure*}

The momentum exchange formula (\ref{delta_p}) may be used to test our code against the LSB Hamiltonian (\ref{SC2-LSBHamiltonian}) for problems outside the scope of Newtonian gravity; in particular, equation (\ref{delta_p}) makes use of $H_2$ (\ref{HamiltonianPart2}) and $H_3$ (\ref{HamiltonianPart3}) in the LSB Hamiltonian, and describes the momentum exchange for the scattering of ultrarelativistic particles.

In Figure \ref{FIG-ScatteringTestsRelativeError}, we present the results of our comparison tests for scattering problems. We plot the following formula for the relative error in the magnitude of the momentum exchange $\Delta p$ for our scattering tests:
\begin{equation} \label{relerror}
\text{Relative Error (\%)} = 100 \times \left| \frac{\Delta p_{num} - \Delta p_{an}}{\Delta p_{an}} \right|
\end{equation}

\noindent where $\Delta p_{an}$ is the magnitude of the momentum exchange given by formula (\ref{delta_p}), and $\Delta p_{num}$ is our numerical result for the magnitude of the momentum exchange. The plots in Figure \ref{FIG-ScatteringTestsRelativeError} show that for $b<10^{10}$, the error in our scattering tests scales as $1/b$, as one might expect. The straight-line approximation used to obtain Equation (\ref{delta_p}) is only valid for high values\footnote{We attribute the large errors $\sim 10 \%$ errors for the lowest values of $b$ in Figure \ref{FIG-ScatteringTestsRelativeError} to the failure of the straight-line approximation used to obtain Equation (\ref{delta_p}).} for $b$, so that the errors may be expanded in $1/b$; it follows that for large $b$, the errors scale as $1/b$. For $b>10^{11}$, the relative error levels off and remains at some value on the order of $10^{-9} \, \%$. We attribute this behavior for $b>10^{11}$ to truncation error. The reader may note that for the massless-massless and massless-massive cases, the relative error becomes extremely small around $b \sim 10^{11}$ before leveling off for $b>10^{11}$; this behavior is due to a change in sign for the quantity $\Delta p_{num} - \Delta p_{an}$ around $b \sim 10^{11}$, which does \textit{not} occur for the massless-massive case.

We would like to also report that our scattering tests were consistent with conservation of momentum. In the two-particle scattering tests we performed, the components of the momenta for each particle remained equal and opposite to the other up to machine precision. Since the scattering tests were formulated in the center-of-mass frame, this demonstrates that $\codename$ conserves momentum in two-particle scattering problems.

%
%
\subsection{Energy Dissipation}
\label{sec_dissipation}

 \begin{figure*}
 \includegraphics[width=\linewidth]{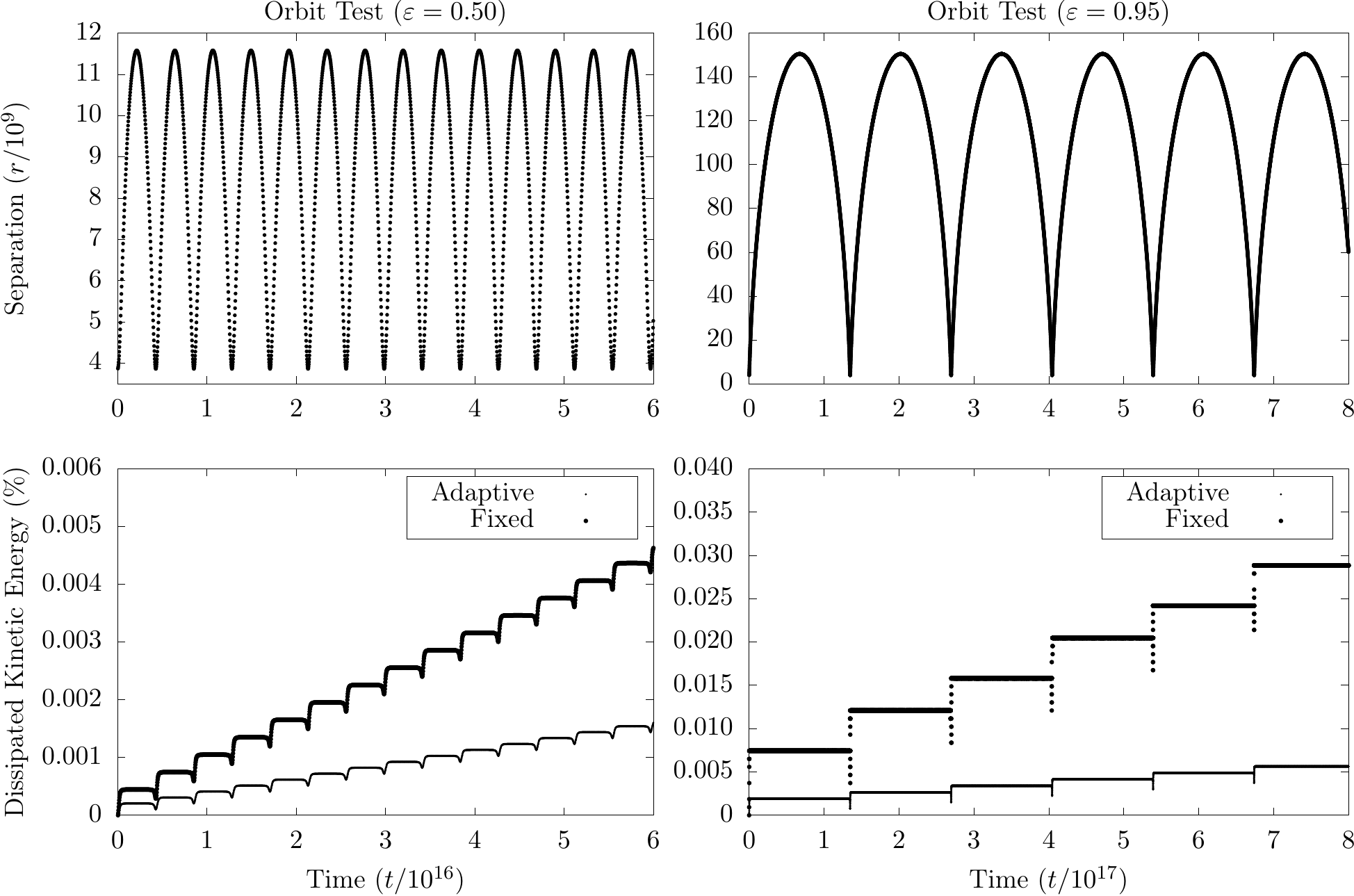}
\caption{Plots of the rescaled separation and the dissipated kinetic energy (energy lost) with respect to time for the orbit test cases \textbf{(a)} and \textbf{(b)} in Figure \ref{FIG-ConvergenceTests}. The dissipated energy plots include both fixed timestep ($h=2.7 \times 10^{13}$) and adaptive timestep cases (with Courant factor of $0.1$). Again, we perform computations to quadruple floating-point precision. Here, the kinetic energy is defined as the Hamiltonian minus the rest masses of the particles in the system, and the percentage dissipated is calculated with respect to the initial value. The rescaled separation parameter $r/10^{9}$ (with $r:=|\textbf{q}_2-\textbf{q}_1|$) has been plotted to show that the most of the dissipation occurs near the point of closest approach (both plots use data from the adaptive timestep runs).}
\label{FIG-EnergyTests}
\end{figure*}

Runge-Kutta algorithms typically exhibit dissipative behavior; since $\codename$ uses an RK4 integrator, one might expect to find dissipative behavior, particularly for long-timescale simulations. In Figure \ref{FIG-EnergyTests}, we plot the dissipated kinetic energy as a percentage of the initial kinetic energy (kinetic energy here being defined as the value of the Hamiltonian minus the sum of the rest masses) for the high-eccentricity orbit test cases \textbf{(a)} and \textbf{(b)} presented in Figure \ref{FIG-ConvergenceTests}. The plots show that $\codename$ does indeed exhibit dissipative behavior for high-eccentricity orbits, and they demonstrate that on average, the dissipated kinetic energy increases linearly with time. Upon comparing the plots for dissipated kinetic energy and the rescaled separation between the particles, we find that most of the dissipation occurs at the point of closest approach. We also note a greater relative dissipation for the orbit with higher eccentricity ($\varepsilon = 0.95$). Figure \ref{FIG-EnergyTests} includes plots for both fixed timestep and adaptive timestep test cases, and we find that the dissipated kinetic energy grows at a decreased rate when adaptive timestepping is used.

\section{Applications and Future Work}
Since $\codename$ is obtained from a PM approximation, it is well suited for modeling any astrophysical problem that involves weak gravitational interactions between $N$ compact objects, some or all of which are ultrarelativistic (fast moving), provided that the timescale for gravitational interactions remains relatively short. As shown by our scattering tests in section \ref{sec_convergence}, successfully modeling light deflection in $\codename$ is simply a matter of assigning one or more of the $N$ bodies to be photons.  With light therefore grouped on equal footing with all of the other particles, $\codename$ makes it a straightforward task to model $N$-body light deflection just like any $N$-body gravitational dynamics problem.  This could prove particularly useful in situations in which we desire a time-dependent solution to a light deflection problem, for example, the deflection of light by a system of gravitationally interacting bodies like a binary system or a planetary system.  

Light deflection by planetary systems has been used to find the first circumbinary planet \cite{circumbinary_planet}.  This is a three-body problem that involves two stars and one planet, although in our code the light itself would constitute a fourth body.  Multiple models are in contention to describe this system, such as a planet orbiting a binary star versus a planet orbiting a single star within a widely separated binary.  $\codename$ would be very well-suited for modeling such a system, as well as more complicated systems with higher $N$ such as the multiple-planet system where a Jupiter analog was detected via light deflection \cite{Jupiter_microlensing}.

Another scenario that could be modeled by the weak-field fast-motion gravitational dynamics of $\codename$ is that of hyper-velocity stars, such as those stars ejected from a galaxy by a binary black hole merger, which can result in stars ejected with a speed of $\frac{1}{3} c$.  \cite{Loeb1} \cite{Loeb2} 

While $\codename$ is in principle an $N$-body code in the sense that that we place no hard limit on the number of particles that $\codename$ can accept, the $O(N^2)$ scaling limits the number of particles that $\codename$ can model in practice. As a result, $\codename$ is at present only suited for problems that have a limited number of particles, but are still too complicated to work out analytically. At the moment, support for parallel processing has not yet been implemented in $\codename$, which presents another limitation on the number of particles $\codename$ can handle; we are currently in the process of implementing parallel processing in $\codename$. 

A current limitation of $\codename$ comes from the dissipative nature of the RK4 integrator, as illustrated in Figure \ref{FIG-EnergyTests}; this renders our present code unsuitable for long-timescale $N$-body simulations. In the future, we intend to implement or add support for integrators that have improved energy conservation on long timescales. The methods under consideriation include implicit time-symmetric Runge-Kutta or symplectic integrators \cite{Hutetal1997,Hutetal1995} (implemented by way of iteration), the partitioned Runge-Kutta method \cite{SanzSernaCalvo1994}, the splitting methods of Tao \cite{Tao2016}, or the implementation of exactly conservative integrators of the type described in \cite{Shadwicketal1995}. 

\begin{acknowledgments}
This work was partially supported by the National Science Foundation under Grant No. PHY-1620610. We thank Mark Selover for his advice and feedback.
\end{acknowledgments}

\bibliography{pomin}

\end{document}